\documentclass{article}
\usepackage{spconf,amsmath,graphicx,amsfonts,epsfig}
\usepackage{times}
\usepackage{amsbsy}
\usepackage{latexsym}
\usepackage{amssymb}

\title{Joint PIC and relay selection based on greedy techniques for cooperative DS-CDMA systems}

\name{Jiaqi Gu{$^*$} and Rodrigo C. de Lamare{$^{*\dag}$}}
\address{Communications Research Group, Department of Electronics,
University of York, U.K.{$^*$}\\CETUC, PUC-Rio, Rio de Janeiro, Brazil{$^\dag$}\\
Email: jg849@york.ac.uk, rodrigo.delamare@york.ac.uk}

\begin{document}
\ninept
\linespread{0.8}
\setlength{\abovedisplayskip}{0.4mm}
\setlength{\belowdisplayskip}{0.4mm}
\setlength\floatsep{7.5pt}
\setlength\textfloatsep{7.5pt}
\setlength\intextsep{7.5pt}
\setlength{\abovecaptionskip}{0.5pt}
\setlength{\belowcaptionskip}{0.5pt}

\maketitle
\let\thefootnote\relax\footnote{This work is funded by the ESII consortium under task 26 for
low-cost wireless ad hoc and sensor networks}\vspace{-2em}

\begin{abstract}
In this work, we propose a cross-layer design strategy based on
the parallel interference cancellation (PIC) detection
technique and a multi-relay selection algorithm for the uplink of
cooperative direct-sequence code-division multiple access (DS-CDMA)
systems. We devise a low-cost greedy list-based PIC (GL-PIC) strategy with RAKE receivers as
the front-end that can approach the maximum likelihood detector
performance. We also present a low-complexity multi-relay selection algorithm
based on greedy techniques that can approach the performance of
an exhaustive search. Simulations show an excellent bit error rate
performance of the proposed detection and relay selection algorithms
as compared to existing techniques.
\end{abstract}

\section{Introduction}
\vspace{-0.5em}
\label{sec:intro}
Multipath fading is a major constraint that seriously limits the
performance of wireless communications. Indeed, severe fading has a detrimental
effect on the received signals and can lead to a degradation of the
transmission of information and the reliability of the network.
Cooperative diversity is a modern technique that has been
widely considered in recent years \cite{Proakis} as an effective tool to
deal with this problem. Several cooperative schemes have been proposed in the
literature \cite{sendonaris,Venturino,laneman04}, and among the most
effective ones are Amplify--and--Forward (AF) and
Decode--and--Forward (DF) \cite{laneman04}.

DS-CDMA systems are a multiple access technique that can be
incorporated with cooperative systems in ad hoc and sensor networks
\cite{Bai,Souryal,Levorato}. Due to the multiple access interference
(MAI) effect that arises from nonorthogonal received waveforms, the
system is adversely affected. To deal with this issue, multiuser
detection (MUD) techniques have been developed \cite{Verdu1} as an
effective approach to suppress MAI. The optimal detector, known as
maximum likelihood (ML) detector, has been proposed in
\cite{Verdu2}. However, this method is infeasible for ad hoc and
sensor networks considering its computational complexity. Motivated
by this fact, several sub-optimal strategies have been developed:
the linear detector \cite{Lupas,jio,jidf,jio_stcdma}, the successive
interference cancellation (SIC) \cite{Patel}, the parallel
interference cancellation (PIC) \cite{Varanasi,itic} and the minimum
mean-square error (MMSE) decision feedback detector
\cite{RCDL1,mbdf}.

In cooperative relaying systems, different strategies that utilize
multiple relays have been recently introduced in
\cite{Jing,Clarke,Ding,Song,jpa_alt,Talwar,armo}. Among these
approaches, a greedy algorithm is an effective way to approach the
global optimal solution. Greedy algorithms have been widely applied
in sparse approximation \cite{Tropp}, internet routing \cite{Flury}
and arithmetic coding \cite{Jia}. In relay assisted systems, greedy
algorithms are used in \cite{Ding,Song} to search for the best relay
combination, however, with insufficient numbers of combinations
considered, a significant performance loss is experienced as
compared to an exhaustive search.

The aim of this work is to propose a cross-layer approach that jointly
considers the optimization of a low-complexity detection and a relay selection algorithm for ad hoc and sensor networks that employ DS-CDMA
systems. Cross-layer designs that integrate different layers of the network have been employed in prior
work \cite{RCDL2,Chen} to guarantee the quality of service and help increase the
capacity, reliability and coverage of systems. However, involving MUD techniques
with relay selection in cooperative relaying systems has not been discussed widely
in the literature. In \cite{Venturino,Cao}, an MMSE-MUD technique has been applied
to cooperative systems, the results indicate that the transmissions are
more resistant to MAI and obtain a significant performance gain when
compared with a single direct transmission. However, extra
complexity is introduced, as matrix inversions are required
when an MMSE filter is deployed.

In this work, we devise a low-cost greedy list-based parallel
interference cancellation (GL-PIC) strategy with RAKE receivers as
the front-end that can approach the maximum likelihood detector
performance. Unlike prior art, the proposed GL-PIC algorithm
exploits the Euclidean distance between users of interest and the
nearest constellation points, re-examines the reliability of the
estimates so that all possible combination lists of tentative decisions
can be checked. With this greedy-like approach, an improved detection
performance can be obtained. We also present a low-complexity
multi-relay selection algorithm based on greedy techniques that can
approach the performance of an exhaustive search. In the proposed
greedy algorithm, a selection rule is employed via several stages.
At each stage, a limited number of relay combinations is examined and compared,
resulting in a low-cost strategy to approach the performance of
an exhaustive search. A cross-layer design strategy that brings
together the proposed GL-PIC algorithm and the greedy relay
selection is then considered and evaluated by computer simulations.

The rest of this paper is organized as follows. In Section 2, the
system model is described. In Section 3, the GL-PIC multiuser
detection method is presented. In Section 4, the relay selection
strategy is proposed. In Section 5, simulation results are
presented and discussed. Finally, conclusions are drawn in Section
6.
\vspace{-1em}
\section{Cooperative DS-CDMA system model}
\vspace{-1em}
\label{sec:sys}
\begin{figure}[!htb]
\begin{center}
\def\epsfsize#1#2{0.8\columnwidth}
\epsfbox{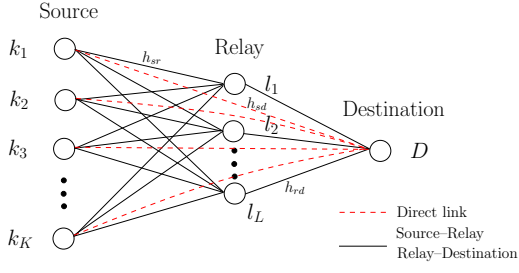} \caption{\footnotesize
Uplink of a cooperative DS-CDMA system.} \vskip -5pt \label{fig1}
\end{center}
\vspace{-1em}
\end{figure}

We consider the uplink of a synchronous DS-CDMA system with $K$
users $(k_1,k_2,...k_K)$, $L$ relays $(l_1,l_2,...l_L)$, $N$ chips per
symbol and $L_p$ $(L_p<N)$ propagation paths for each link. The system
is equipped with a DF protocol at each relay and we assume that the
transmit data are organized in packets comprising $P$ symbols. The
received signals are filtered by a matched filter, sampled at chip
rate to obtain sufficient statistics and organized into $M \times1$
vectors $\textbf{y}_{sd}$, $\textbf{y}_{sr}$ and $\textbf{y}_{rd}$,
which represent the signals received from the sources (users) to the
destination, the sources to the relays and the relays to the
destination, respectively. The proposed algorithms for interference
mitigation and relay selection are employed at the relays and at the
destination. As shown in Fig.\ref{fig1}, the received signal at the
destination comprises the data transmitted during two phases that are
jointly processed at the destination. Therefore, the received signal is described by a
$2M\times1$ vector formed by stacking the received signals
from the relays and the sources as given by
\begin{equation}
\begin{split}
\hspace{-0.5em} \left[\hspace{-0.5em}\begin{array}{l}
  \textbf{y}_{sd}\\
  \textbf{y}_{rd}\\
\end{array} \hspace{-0.5em} \right] & = \left[\hspace{-0.5em} \begin{array}{l}
  \sum\limits_{k=1}^K  a_{sd}^k\textbf{S}_k\textbf{h}_{sd,k}b_k \\
  \sum\limits_{l=1}^{L}\sum\limits_{k=1}^K a_{r_ld}^k\textbf{S}_k\textbf{h}_{r_ld,k}\hat{b}_{r_ld,k}\\
\end{array}\hspace{-0.5em} \right] + \left[ \hspace{-0.5em} \begin{array}{l}
  \textbf{n}_{sd}\\
  \textbf{n}_{rd}\\
  \end{array}\right], \label{equation1}
\end{split}
\end{equation}
where $M=N+L_p-1$, $b_k\in\{+1,-1\}$ correspond to the transmitted
symbols, $a_{sd}^k$ and $a_{r_ld}^k$ represent the $k$-th user's amplitude from
the source to the destination and from the $l$-th relay to the destination.
The $M \times L_p$ matrix $\textbf{S}_k$ contains the signature sequence of each
user shifted down by one position at each column that forms
\begin{equation}
\textbf{S}_k = \left[\begin{array}{c c c }
s_{k}(1) &  & {\bf 0} \\
\vdots & \ddots & s_{k}(1)  \\
s_{k}(N) &  & \vdots \\
{\bf 0} & \ddots & s_{k}(N)  \\
 \end{array}\right],
\end{equation}
where $\textbf{s}_k=[s_k(1),s_k(2),...s_k(N)]^T$ is the signature
sequence for user $k$. The vectors
$\textbf{h}_{sd,k}$, $\textbf{h}_{r_ld,k}$ are the $L_p\times1$
channel vectors for user $k$ from the source to the destination and the $l$-th
relay to the destination. The $M\times1$ noise vectors
$\textbf{n}_{sd}$ and $\textbf{n}_{rd}$ contain samples of zero
mean complex Gaussian noise with variance $\sigma^2$, $\hat{b}_{r_ld,k}$
is the decoded symbol at the output of relay $l$ after using the DF protocol.
The received signal in (\ref{equation1}) can then be described by
\begin{equation}
\textbf{y}_d(i)=\sum\limits_{k=1}^K \textbf{C}_k
\textbf{H}_k(i)\textbf{A}_k(i)\textbf{B}_k(i)+\textbf{n}(i),
\end{equation}
where $i$ denotes the time instant corresponding to one symbol in the
transmitted packet and its received and relayed copies. $\textbf{C}_k$
is a $2M\times(L+1)L_p$ matrix comprising shifted versions of $\textbf{S}_k$ as given by
\begin{equation}
\textbf{C}_k = \left[\begin{array}{c c c c}
\textbf{S}_{k} & {\bf 0} & \ldots & {\bf 0} \\
{\bf 0} & \ \ \textbf{S}_{k} & \ldots & \ \ \textbf{S}_{k}\\
 \end{array}\right],
\end{equation}
$\textbf{H}_k(i)$ represents a $(L+1)L_p \times (L+1)$ channel matrix
between the sources and the destination and the relays and the
destination links. $\textbf{A}_k(i)$ is a $(L+1)\times(L+1)$
diagonal matrix of amplitudes for user $k$. The matrix
$\textbf{B}_k(i)=[b_k,\hat{b}_{r_1d,k},\hat{b}_{r_2d,k},...\hat{b}_{r_Ld,k}]^T$
is a $(L+1)\times1$ matrix for user $k$ that contains the transmitted symbol at
the source and the detected symbols at the output of each relay, and
$\textbf{n}(i)$ is a $2M\times1$ noise vector.
\vspace{-1em}
\section{The proposed GL-PIC multiuser detector}
\vspace{-1em}
\label{sec:PIC}
\begin{figure}[!htb]
\begin{center}
\def\epsfsize#1#2{1\columnwidth}
\epsfbox{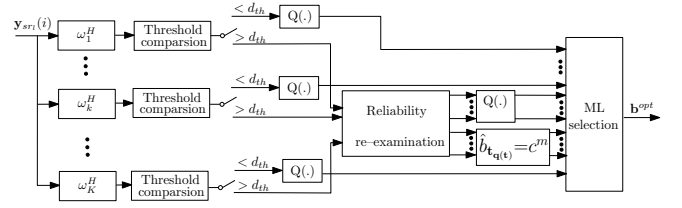} \caption{\footnotesize
\vspace{-2em}
Block diagram of the proposed GL-PIC multi-user detector.} \vskip -5pt \label{fig2}
\end{center}
\vspace{-1em}
\end{figure}

\begin{figure}[!htb]
\vspace{-0.8em}
\begin{center}
\def\epsfsize#1#2{1\columnwidth}
\epsfbox{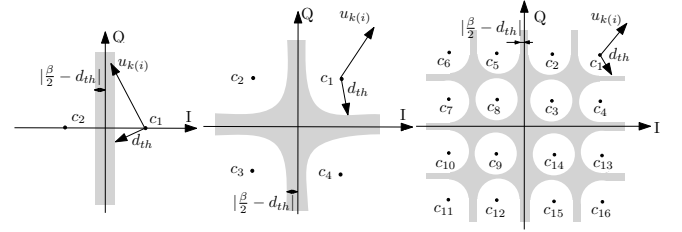} \vspace{-2.5em}\caption{\footnotesize
The threshold comparison for reliability check in BPSK, QPSK and 16 QAM constellations.}
\vspace{-2em}
\label{fig3}
\end{center}
\end{figure}

In this section, we present a GL-PIC detector
that can be applied at both the relays and destination in the
uplink of a cooperative system. The GL-PIC detector uses the
RAKE receiver as the front-end, which reduces computational
complexity by avoiding the matrix inversion required when
MMSE filters are applied. With the structure depicted in
Fig.\ref{fig2}, the proposed GL-PIC algorithm determines the reliability
of the detected symbol by comparing the Euclidean distance between the
symbol of users of interest and the potential nearest constellation point
with a chosen threshold. After checking the reliability of the symbol
estimates by listing all possible combinations of tentative
decisions, the $n_q$ most unreliable users are re-examined
via a number of selected constellation points in a greedy-like approach,
which saves computational complexity by avoiding redundant processing
with reliable users. Following the diagram in
Fig.\ref{fig2}, the soft estimates of the RAKE receiver for each
user are obtained by
\begin{equation}
u_{k}(i)=\textbf{w}_{k}^{H}\textbf{y}_{sr_l}(i),
\end{equation}
where $\textbf{y}_{sr_l}(i)$ represents the received signal
from the source to the $l$-th relay, $u_{k}(i)$ stands for
the soft output of the $i$-th symbol for user $k$ and $\textbf{w}_{k}^{H}$
denotes the RAKE receiver that corresponds to a filter matched to
the effective signature at the receiver. As shown by Fig.\ref{fig3},
$\beta$ is the distance between two nearest constellation
points, $d_{th}$ is the threshold. For the $k$-th user,
the reliability of its soft estimates is determined by the
Euclidean distance between $u_{k}(i)$ and its nearest constellation
points $c$.
\\
\textbf{Decision reliable}:\\
If the soft estimation of $n_a$ users satisfy the following condition
\begin{equation}
u_{\textbf{t}_a(t)}(i)\notin \textbf{C}_{\rm grey},\ \ \ {\rm for} \ t\in [1,2,...,n_a],
\end{equation}
where $\textbf{t}_a$ is a $1\times n_a$ vector that contains $n_a$ reliable estimates, $\textbf{C}_{\rm grey}$ is the grey area in Fig.\ref{fig3}
and the grey area would extend unlimitedly along both the vertical and horizontal directions. These soft estimates will be applied to a slicer $Q(\cdot)$ as
described by
\begin{equation}
\hat{b}_{\textbf{t}_a(t)}(i)=Q(u_{\textbf{t}_a(t)}(i)),\ {\rm for}\ t\in [1,2,...,n_a],
\end{equation}
where $\hat{b}_{\textbf{t}_a(t)}(i)$ denotes the detected symbol for the $\textbf{t}_a(t)$-th user.\\
\textbf{Decision unreliable}:\\
In case that $n_b$ users are determined as unreliable, a $1\times n_b$
vector $\textbf{t}_b$ with $n_b$ unreliable estimates included is produced, as given by
\begin{equation}
u_{\textbf{t}_b(t)}(i)\in \textbf{C}_{\rm grey},\ \ \ {\rm for} \ t\in [1,2,...,n_b],
\end{equation}
we then sort these unreliable estimates in terms of their Euclidean
distance in a descending order. Consequently, the first $n_q$ users from the ordered set are deemed as the most
unreliable ones as they experience the farthest distance to their reference constellation
points. These $n_q$ estimates are then examined in terms of all possible constellation values
$c^m$ $(m=1,2,...,N_c)$ from the $1\times N_c$ constellation points set $\textbf{C}\subseteq \textsl{F}$, where $\textsl{F}$ is a subset of the complex field, and $N_c$ is determined by the modulation type. Meanwhile, the remaining $n_p=n_b-n_q$ unreliable
users are applied to the slicer $Q(\cdot)$ directly, as described by
\begin{equation}
\hat{b}_{\textbf{t}_p(t)}(i)=Q(u_{\textbf{t}_p(t)}(i)),
\ \ {\rm for} \ t\in [1,2,...,n_p],
\end{equation}
\begin{equation}
\hat{b}_{\textbf{t}_q(t)}(i)=c^m,\ \
{\rm for} \ t\in [1,2,...,n_q],
\vspace{-0.4em}
\end{equation}
where $\textbf{t}_p\cap \textbf{t}_q=\varnothing$ and $\textbf{t}_p \cup \textbf{t}_q=\textbf{t}_b$.
Therefore, by listing all possible combinations of elements across the $n_q$ most unreliable
users, the following $K\times1$ tentative candidate decision lists are generated
\begin{equation}
\textbf{b}^j=[\textbf{s}_{a},\ \ \textbf{s}_{p},\ \ \textbf{s}^j_{q}]^T, \ j=1,2,...,N_c^{n_q},
\end{equation}
where $\textbf{s}_{a} \ = \ [\hat{b}_{\textbf{t}_a(1)},\hat{b}_{\textbf{t}_a(2)},...,\hat{b}_{\textbf{t}_a(n_a)}]^T$
is a $n_a \times 1$ vector that contains the detected values for the $n_a$ reliable users,
$\textbf{s}_{p}=[\hat{b}_{\textbf{t}_p(1)},\hat{b}_{\textbf{t}_p(2)},...,\hat{b}_{\textbf{t}_p(n_p)}]^T$ is a
$n_p \times 1$ vector that represents $n_p$ unreliable users that are detected by the slicer $Q(\cdot)$ directly,
and $\textbf{s}^j_q=[c_{\textbf{t}_q(1)}^{m},c_{\textbf{t}_q(2)}^{m},...,c_{\textbf{t}_q(n_q)}^{m}]^T$ is a $n_q \times 1$
tentative candidate combination vector. Each entry of the vector is selected randomly from the constellation
point set $\textbf{C}$ and all possible $N_c^{n_q}$ combinations need to be considered and examined.
The trade-off between performance and complexity is highly related to the
the modulation type and the number($n_q$) of users we choose from $\textbf{t}_b$. Additionally,
the threshold we set at the initial stage is also a key factor that could affect the quality of detection.

After the $N_c^{n_q}$ candidate lists are generated, the ML rule is used subsequently to choose the best candidate list
as described by
\begin{equation}
\textbf{b}^{\textrm{opt}}= \min _{\substack{1\leq j\leq N_c^{n_q}}}\mid \textbf{y}_{sr_l}(i)-\textbf{H}_{sr_l}\textbf{b}^{j}(i)\mid^2.
\end{equation}
Following that, $\textbf{b}^{\textrm{opt}}$ is used as the input for a multi-iteration PIC process as described by
\begin{equation}
\hat{b}_k^i=Q(\textbf{H}_{sr_l,k}^H\textbf{y}_{sr_l}-\sum\limits_{\substack{j=1\\j\neq k}}^{K}\textbf{H}_{sr_l,k}^H\textbf{H}_{sr_l,j}\hat{b}_j^{i-1}),
\vspace{-0.5em}
\end{equation}
where $\hat{b}_k^i$ denotes the detected value for user $k$ at the $i$-th PIC iteration, $\textbf{H}_{sr_l,k}$ and $\textbf{H}_{sr_l,j}$
stand for the channel matrices for the $k$-th and $j$-th user from the source to the $l$-th relay, respectively. $\hat{b}_j^{i-1}$
is the detected value for user $j$ that comes from the $(i-1)$-th PIC iteration. Normally, the conventional PIC is performed in a multi-iteration way,
where for each iteration, PIC simultaneously subtracts off the interference for each user produced by the remaining ones. The MAI generated
by other users is reconstructed based on the tentative decisions from the previous iteration. Therefore, the accuracy
of the first iteration would highly affect the PIC performance as error propagation occurs when
incorrect information imports. In this case, with the help of the GL-PIC algorithm, we are able
to improve the accuracy of the detection and obtain better performance.
\vspace{-1em}
\section{proposed greedy multi-relay selection method}
\vspace{-1.2em}
\label{sec:relay}
In this section, a greedy multi-relay selection method is introduced. For this problem, an exhaustive search of all possible subsets of relays is needed to attain the optimum relay combination. However, an exhaustive search presents a considerable computational complexity, limiting its application in practical systems. With $L$ relays involved in the transmission, an exponential complexity of $2^L-1$ would be required. This fact motivates us to seek other alternative methods. By mitigating the poorest relay-destination link stage by stage, the standard greedy algorithm can be used in the selection process, yet only a local optimum can be achieved. Unlike the traditional ways, the proposed greedy multi-relay selection method can go through a sufficient number of relay combinations and approach the best one based on previous decisions. In the proposed relay selection, the signal to interference and noise ratio (SINR) is used as the criterion to determine the optimum relay set. The expression of the SINR is expressed by
\begin{equation}
 {\rm SINR_q} =\frac{E[|\textbf{w}_q^H\textbf{r}|^2]}{E[|\boldsymbol\eta|^2]+\textbf{n}},
\end{equation}
where $\textbf{w}_q$ denotes the RAKE receiver for user $q$, $E[|\boldsymbol\eta|^2]=E[|\sum\limits_{\substack{k=1\\k\neq q}}^K\textbf{H}_kb_k|^2]$ is the interference brought by all other users, $\textbf{n}$ is the noise vector. For the RAKE receiver, the SINR is given by
\begin{equation}
 {\rm SINR_q} =\frac{\textbf{H}_q^H\textbf{H}\textbf{H}^H\textbf{H}_q}{\textbf{H}_\eta \textbf{H}_\eta^H+\textbf{H}_q^H\sigma_N^2\textbf{H}_q},
\end{equation}
where $\textbf{H}_q$ is the channel matrix for user $q$, $\textbf{H}$ is the channel matrix for all users, $\textbf{H}_\eta$ represents the channel matrix of all other users except user $q$. It should be mentioned that in various relay combinations, the channel matrix $\textbf{H}_q$ for user $q$ $(q=1,2,...,K)$ is different as different relay-destination links are involved, $\sigma_N^2$ is the noise variance. This problem thus can be cast as the following optimization:
\begin{equation}
 {\rm SINR_{\Omega_{best}}}= max \ \ \{min({\rm SINR_{\Omega_{r(q)}}}), q=1,...,K\},
\end{equation}
where $\Omega_{r}$ denotes all possible combination sets $(r \leq L(L+1)/2)$ of any number of selected relays, ${\rm SINR_{\Omega_{r(q)}}}$ represents the SINR for user $q$ in set $\Omega_r$, min $({\rm SINR_{\Omega_{r(q)}}}) = {\rm SINR_{\Omega_r}}$ stands for the SINR for relay set $\Omega_r$ and
$\Omega_{\rm best}$ is the best relay set that provides the highest SINR.
\vspace{-1.3em}
\subsection{Standard greedy relay selection algorithm}
\vspace{-0.7em}
The standard greedy relay selection method works in stages by removing the single relay according to the channel path power, as given by
\begin{equation}
 P_{h_{r_ld}}=\textbf{h}_{r_ld}^H\textbf{h}_{r_ld},\vspace{-0.5em}
\end{equation}
where $\textbf{h}_{r_ld}$ is the channel vector between the $l$-th relay and the destination. At the first stage, the initial SINR is determined
when all $L$ relays are involved in the transmission. Consequently, we cancel the worst relay-destination link and calculate the current SINR for the remaining $L-1$ relays, as compared with the previous SINR, if
\begin{equation}
{\rm SINR_{cur}}>{\rm SINR_{pre}}, \label{equation6}
\end{equation}
we update the previous SINR as
\begin{equation}
{\rm SINR_{pre}} = {\rm SINR_{cur}}, \label{equation7}
\end{equation}
and move to the third stage by removing the current poorest link and repeating the above process. The algorithm stops either when ${\rm SINR_{cur}}<{\rm SINR_{pre}}$ or when there is only one relay left. The selection can be performed once at the beginning of each packet transmission.
\vspace{-1.2em}
\subsection{Proposed greedy relay selection algorithm}
\vspace{-0.7em}
In order to improve the performance of the standard algorithm, we propose a new greedy relay selection algorithm that is able to achieve a good balance between the performance and the complexity. The proposed method differs from the standard technique as we drop each of the relays in turns rather than drop them based on the channel condition at each stage. The algorithm can be summarized as:
\begin{enumerate}
\item Initially, a set $\Omega_A$ that includes all $L$ relays is generated and its corresponding SINR is calculated, denoted by ${\rm SINR_{pre}}$.
\item For the second stage, we calculate the SINR for $L$ combination sets with each dropping one of the relays from $\Omega_A$. After that, we pick the combination set with the highest SINR for this stage, recorded as ${\rm SINR_{cur}}$.
\item Compare ${\rm SINR_{cur}}$ with the previous stage ${\rm SINR_{pre}}$, if (\ref{equation6}) is true, we save this corresponding relay combination as $\Omega_{\textrm{cur}}$ at this stage. Meanwhile, we update the ${\rm SINR_{pre}}$ as in (\ref{equation7}).
\item After moving to the third stage, we drop relays in turn again from $\Omega_{\textrm{cur}}$ obtained in stage two. $L-1$ new combination sets are generated, we then select the set with the highest SINR and repeat the above process in the following stages until either ${\rm SINR_{cur}}<{\rm SINR_{pre}}$ or there is only one relay left.
\end{enumerate}
This new greedy selection method considers the combination effect of the channel condition so that additional useful sets are examined. When compared with the standard greedy relay selection method, the previous stage decision is more accurate and the global optimum can be approached more closely. Furthermore, its complexity is less than $L(L+1)/2$, which is much lower than the exhaustive search. Similarly, the whole process is performed only once before each packet.
\vspace{-1em}
\begin{table}[!htp]\scriptsize
\footnotesize
\centering\caption{The proposed greedy multi-relay selection algorithm}
\begin{tabular}{l}
\hline
$\Omega_A=[1,2,3,...L]$\% $\Omega_A$ denotes the set when all relays are involved\\
${\rm SINR_{\Omega_A}}=\textrm{min}({\rm SINR_{\Omega_{A(q)}}}), q=1,2,...K$\\
${\rm SINR_{pre}}={\rm SINR_{\Omega_A}}$\\
\textbf{for} stage =1 \textbf{to} $L-1$\\ \ \ \ \ \
\textbf{for} $r$=1 \textbf{to} $L+1$-stage\\ \ \ \ \ \ \ \ \ \
$\Omega_r=\Omega_A-\Omega_{A(r)}$\% drop each of the relays in turns\\ \ \ \ \ \ \ \ \ \
${\rm SINR_{\Omega_r}}=\textrm{min}({\rm SINR_{\Omega_r(q)}}), q=1,2,...,K$\\\ \ \ \ \
\textbf{end for}\\ \ \ \ \ \
${\rm SINR_{\textrm{cur}}}=\textrm{max}({\rm SINR_{\Omega_r}})$\\ \ \ \ \ \
$\Omega_{\textrm{cur}}=\Omega_{{\rm SINR_{cur}}}$\\ \ \ \ \ \
\textbf{if} ${\rm SINR_{cur}}>{\rm SINR_{pre}}$ \textbf{and} $\rm{length}(\Omega_{\textrm{cur}})>1$\\ \ \ \ \ \ \ \ \ \
$\Omega_A=\Omega_{\textrm{cur}}$\\ \ \ \ \ \ \ \ \
${\rm SINR_{pre}}={\rm SINR_{cur}}$\\ \ \ \ \ \
\textbf{else}\\ \ \ \ \ \ \ \ \ \ \
\textbf{break}\\ \ \ \ \
\textbf{end if}\\
\textbf{end for}\\
\hline
\end{tabular}
\vspace{-1em}
\end{table}
\vspace{-1em}
\section{simulations}
\vspace{-1em}
\label{sec:siml}
In this section, a simulation study of the proposed GL-PIC multiuser detection strategy with a RAKE receiver and the low cost greedy multi-relay selection method is carried out. The DS-CDMA network uses randomly generated spreading codes of length $N=16$ and employs $L_p=3$ independent paths with the power profile $[0\rm dB,-3\rm dB,-6\rm dB]$ for each source-relay, source-destination and relay-destination link. Their corresponding channel coefficients are taken as uniformly random variables and normalized to ensure the total power is unity. We assume perfectly known channels at the receiver. Equal power allocation with normalization is assumed to guarantee no extra power is introduced during the transmission.  We consider packets with 1000 BPSK symbols and average the curves over 300 trials. For the purpose of simplicity, in the GL-PIC algorithm, a three-iteration PIC process is adopted, $d_{th}=0.25$ and BPSK modulation technique are applied in the following simulations.

\begin{figure} \centerline
{\includegraphics[width=0.85\columnwidth,height=0.675\columnwidth,draft=false]{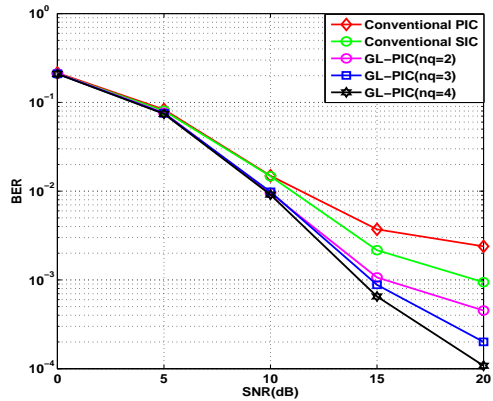}}
\vspace{-1.0em} \caption{BER versus SNR for uplink cooperative
system with different filters employed in the relays and the
destination} \vspace{-1em} \label{fig4}
\end{figure}

\begin{figure}[!htb]
\centerline{
\includegraphics[width=0.45\columnwidth,height=0.675\columnwidth]{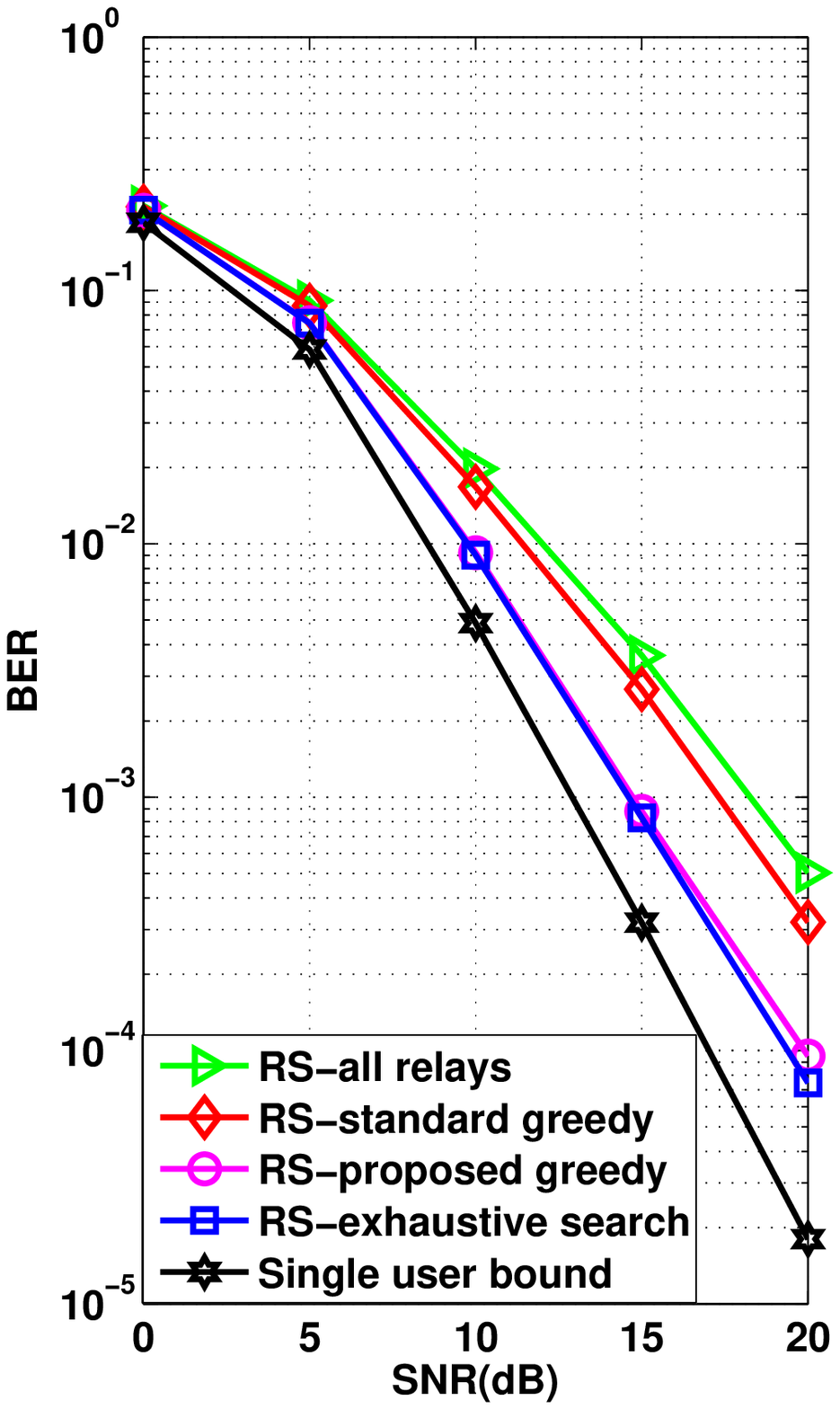}
\includegraphics[width=0.45\columnwidth,height=0.675\columnwidth]{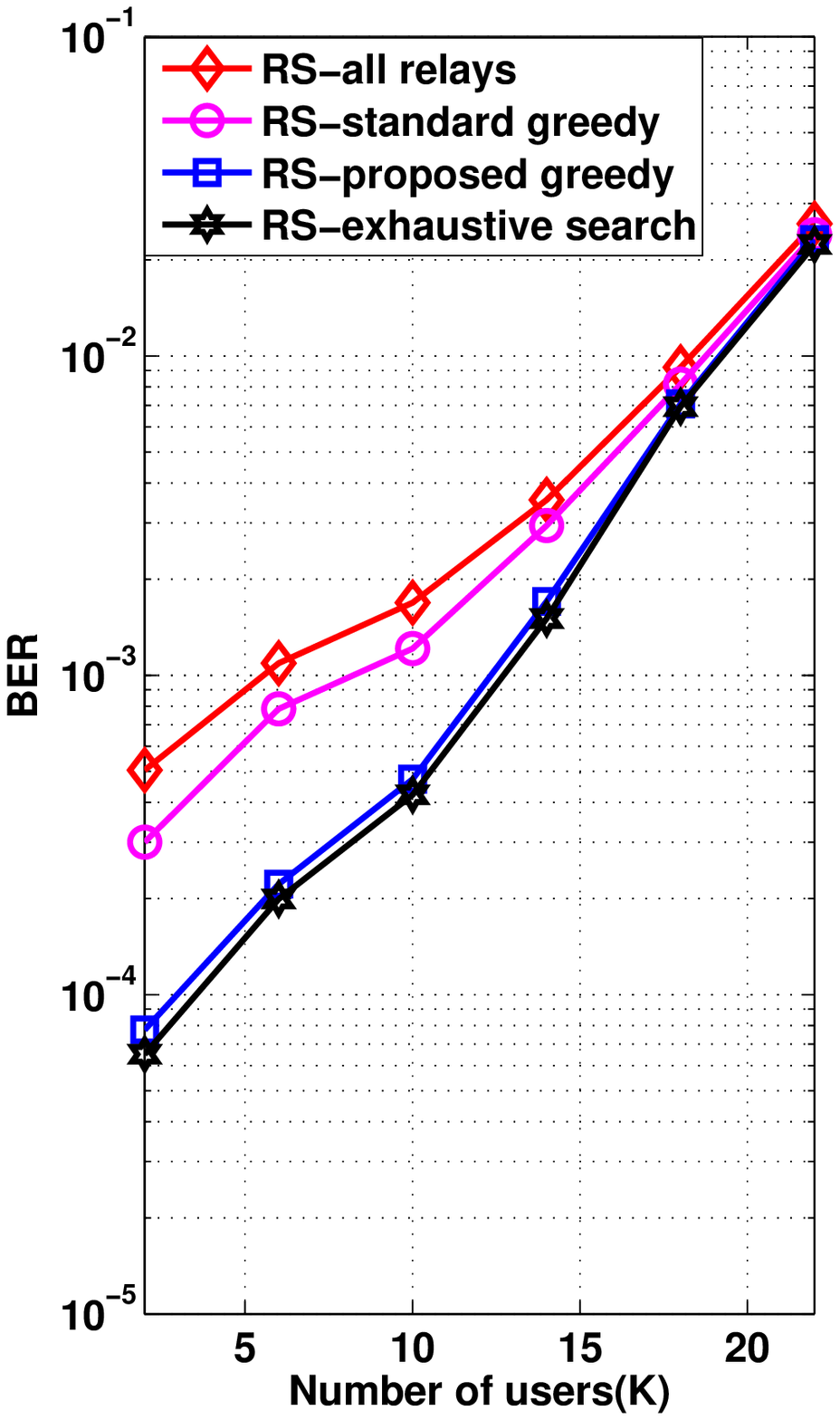}
} \vspace{-1.0em}\caption{\footnotesize a)\ BER versus SNR for
uplink cooperative system\ \ \ b)\ BER versus number of users for
uplink cooperative system} \vspace{-0.2em} \label{fig5}
\end{figure}

The first example, shown in Fig.\ref{fig4} depicts the performance for the proposed cross-layer design, where we compare the effect of different detectors with 10 users and 6 relays when the new greedy multi-relay selection algorithm is applied in the system. Simulation results indicate that the GL-PIC approach allows a more effective reduction of BER, followed by the conventional SIC detector and the conventional PIC detector. Additionally, it is worth noting that some extra performance gains are attained as more $n_q$ unreliable users are selected and re-examined.

The second scenario illustrated in Fig.\ref{fig5}(a) shows the BER versus SNR plot for employing different multi-relay selection strategies, where we apply the GL-PIC detection scheme at both the relays and destination in an uplink cooperative scenario with 10 users and 6 relays. The performance bound for an exhaustive search is presented here for comparison purposes, where it examines all possible relay combinations and picks the one with the highest SINR. From the results, it can be seen that with relay selection, the BER performance substantially improves. Furthermore, the BER performance curve of our proposed multi-relay selection algorithm outperforms the standard greedy algorithm and approaches the same level of the exhaustive search, whilst keeping the complexity reasonably low for practical utilization. The algorithms are then assessed in terms of the BER versus number of users in Fig.\ref{fig5}(b) with a fixed SNR=15dB. Similarly, we apply the GL-PIC detector at both the relays and destination. The results indicate that the overall system performance degrades as the number of users increases. It also suggests that our proposed greedy relay selection method has a big advantage for situations without a high load when compared with the standard greedy relay selection and non-relay selection scenario.

\vspace{-0.5em}
\section{conclusions}
\vspace{-1em}
\label{sec:conc}
A novel cross-layer design strategy that incorporates the greedy list-based parallel interference cancellation (GL-PIC) detection technique and a low cost greedy multi-relay selection algorithm for the uplink of cooperative DS-CDMA systems has been presented in this paper. This approach effectively mitigates the phenomenon of error propagation and selects the optimum relay combination while requiring a low complexity. Simulation results demonstrate that the proposed cross-layer optimization technique can offer considerable gains as compared to existing detectors and can approach the exhaustive search bound very closely.

{\footnotesize
\bibliographystyle{IEEEbib}
\bibliography{reference}}

\begin{thebibliography}{10}

\bibitem{Proakis}
J.G.Proakis,
\newblock {\em Digital Communications},
\newblock McGraw-Hill,Inc, New York, USA, 4th edition, 2011.

\bibitem{sendonaris}
A.~Sendonaris, E.~Erkip, and B.~Aazhang,
\newblock ``User cooperation diversity - parts {I} and {II},''
\newblock {\em IEEE Trans. Communication.}, vol. 51, no. 11, pp. 1927--1948,
  November 2003.

\bibitem{Venturino}
L.~Venturino, X.~Wang, and M.~Lops,
\newblock ``Multi-user detection for cooperative networks and performance
  analysis,''
\newblock {\em IEEE Trans. Signal Processing}, vol. 54, no. 9, pp. 3315--3329,
  September 2006.

\bibitem{laneman04}
J.~N. Laneman and G.~W. Wornell,
\newblock ``Cooperative diversity in wireless networks: Efficient protocols and
  outage behaviour,''
\newblock {\em IEEE Trans. Inf. Theory}, vol. 50, no. 12, pp. 3062--3080,
  December 2004.

\bibitem{Bai}
L.~Bao, L.~Zhao, and Z.~liao,
\newblock ``A novel cooperation scheme in wireless sensor networks,''
\newblock {\em IEEE Wireless Communications and Networking Conference}, pp.
  1889--1893, Las Vegas, NV, Apr. 2008.

\bibitem{Souryal}
M.~R. Souryal, B.~R. Vojcic, and R.~L. Pickholtz,
\newblock ``Adaptive modulation in ad hoc ds/cdma packet radio networks,''
\newblock {\em IEEE Trans. Communication}, vol. 54, no. 4, pp. 714--725, Apr.
  2006.

\bibitem{Levorato}
M.~Levorato, S.~Tomasin, and M.~Zorzi,
\newblock ``Cooperative spatial multiplexing for ad hoc networks with hybrid
  arq: System design and performance analysis,''
\newblock {\em IEEE Trans. Communication}, vol. 56, no. 9, pp. 1545--1555, Sep.
  2008.

\bibitem{Verdu1}
S.Verdu,
\newblock {\em Multiuser Detection},
\newblock Cambridge, 1998.

\bibitem{Verdu2}
S.~Verdu,
\newblock ``Minimum probability of error for asynchronous gaussian
  multiple-access channels,''
\newblock {\em IEEE Trans. Inform. Theory}, vol. IT32, no. 1, pp. 85--96, Jan.
  1986.

\bibitem{Lupas}
R.~Lupas and S.~Verdu,
\newblock ``Linear multiuser detectors for synchronous code-division
  multiple-access channels,''
\newblock {\em IEEE Trans. Inform. Theory}, vol. 35, no. 1, pp. 123--136, Jan.
  1989.

\bibitem{jio}
R.~C. de~Lamare and R.~Sampaio-Neto,
\newblock ``Reduced-rank adaptive filtering based on joint iterative
  optimization of adaptive filters,''
\newblock {\em IEEE Sig. Proc. Letters}, vol. 14, no. 12, pp. 980--983, Dec.
  2007.

\bibitem{jidf}
R.~C. de~Lamare and R.~Sampaio-Neto,
\newblock ``Adaptive reduced-rank processing based on joint and iterative
  interpolation, decimation, and filtering,''
\newblock {\em IEEE Trans. Sig. Proc.}, vol. 57, no. 7, pp. 2503--2514, July
  2009.

\bibitem{jio_stcdma}
R.~C. de~Lamare and R.~Sampaio-Neto,
\newblock ``Reduced-rank space-time adaptive interference suppression with
  joint iterative least squares algorithms for spread-spectrum systems,''
\newblock {\em IEEE Transactions on Vehicular Technology}, vol. 59, no. 3, pp.
  1217--1228, March 2010.

\bibitem{Patel}
P.~Patel and J.~Holtzman,
\newblock ``Analysis of a simple successive interference cancellation scheme in
  ds/cdma systems,''
\newblock {\em IEEE J. Select. Areas Commun.}, vol. 12, no. 5, pp. 796--807,
  Jun. 1994.

\bibitem{Varanasi}
M.~K. Varanasi and B.~Aazhang,
\newblock ``Multistage detection in asynchronous code-division multiple-access
  communications,''
\newblock {\em IEEE Trans. Communication}, vol. 38, no. 4, pp. 509--519, Apr.
  1990.

\bibitem{itic}
R.C. de~Lamare, R.~Sampaio-Neto, and A.~Hjorungnes,
\newblock ``Joint iterative interference cancellation and parameter estimation
  for cdma systems,''
\newblock {\em IEEE Communications Letters}, vol. 11, no. 12, pp. 916--918,
  December 2007.

\bibitem{RCDL1}
R.~C. de~Lamare and R.~Sampaio-Neto,
\newblock ``Minimum mean-squared error iterative successive parallel arbitrated
  decision feedback detectors for ds-cdma systems,''
\newblock {\em IEEE Trans. Communication}, vol. 56, no. 5, pp. 778--789, May.
  2008.

\bibitem{mbdf}
R.C. de~Lamare,
\newblock ``Adaptive and iterative multi-branch mmse decision feedback
  detection algorithms for multi-antenna systems,''
\newblock {\em IEEE Transactions on Wireless Communications}, vol. 12, no. 10,
  pp. 5294--5308, October 2013.

\bibitem{Jing}
Y.~Jing and H.~Jafarkhani,
\newblock ``Single and multiple relay selection schemes and their achievable
  diversity orders,''
\newblock {\em IEEE Trans. Wireless Commun.}, vol. 8, no. 3, pp. 1084--1098,
  Mar 2009.

\bibitem{Clarke}
P.~Clarke and R.~C. de~Lamare,
\newblock ``Transmit diversity and relay selection algorithms for multirelay
  cooperative mimo systems,''
\newblock {\em IEEE Trans. Veh. Technol}, vol. 61, no. 3, pp. 1084--1098, Mar
  2012.

\bibitem{Ding}
M.~Ding, S.~Liu, H.~Luo, and W.~Chen,
\newblock ``{MMSE} based greedy antenna selection scheme for af mimo relay
  systems,''
\newblock {\em IEEE Signal Process. Lett}, vol. 17, no. 5, pp. 433--436, May
  2010.

\bibitem{Song}
S.~Song and W.~Chen,
\newblock ``{MMSE} based greedy eigenmode selection for af mimo relay
  channels,''
\newblock {\em IEEE Globecom}, Anaheim, CA, Dec. 2012.

\bibitem{jpa_alt}
T.~Wang, R.C. de~Lamare, and A.~Schmeink,
\newblock ``Joint linear receiver design and power allocation using alternating
  optimization algorithms for wireless sensor networks,''
\newblock {\em IEEE Transactions on Vehicular Technology}, vol. 61, no. 9, pp.
  4129--4141, Nov 2012.

\bibitem{Talwar}
S.~Talwar, Y.~Jing, and S.~Shahbazpanahi,
\newblock ``Joint relay selection and power allocation for two-way relay
  networks,''
\newblock {\em IEEE Signal Process. Lett}, vol. 18, no. 2, pp. 91--94, Feb
  2011.

\bibitem{armo}
Peng T., R.C. de~Lamare, and A.~Schmeink,
\newblock ``Adaptive distributed space-time coding based on adjustable code
  matrices for cooperative mimo relaying systems,''
\newblock {\em IEEE Transactions on Communications}, vol. 61, no. 7, pp.
  2692--2703, July 2013.

\bibitem{Tropp}
J.~Tropp,
\newblock ``Greedy is good: Algorithmic results for sparse approximation,''
\newblock {\em IEEE Trans. Inf. Theory}, vol. 50, no. 10, pp. 2231–--2242, Oct.
  2004.

\bibitem{Flury}
R.~Flury, S.~V. Pemmaraju, and R.~Wattenhofer,
\newblock ``Greedy routing with bounded stretch,''
\newblock {\em IEEE Infocom.}, Rio de Janeiro, Brazil, Apr. 2009.

\bibitem{Jia}
Y.~Jia, E.~Yang, D.~He, and S.~Chan,
\newblock ``A greedy re-normalization method for arithmetic coding,''
\newblock {\em IEEE Trans. Communication}, vol. 55, no. 8, pp. 1494--–1503,
  Aug. 2007.

\bibitem{RCDL2}
R.~C. de~Lamare,
\newblock ``Joint iterative power allocation and linear interference
  suppression algorithms for cooperative ds-cdma networks,''
\newblock {\em IET, Communications}, vol. 6, no. 13, pp. 1930--1942, Sep. 2012.

\bibitem{Chen}
W.~Chen, L.~Dai, K.~B. Letaief, and Z.~Cao,
\newblock ``A unified cross-layer framework for resource allocation in
  cooperative networks,''
\newblock {\em IEEE Trans. Wireless Commun.}, vol. 7, no. 8, pp. 3000--3012,
  Aug. 2008.

\bibitem{Cao}
Y.~Cao and B.~Vojcic,
\newblock ``{MMSE} multiuser detection for cooperative diversity cdma
  systems,''
\newblock {\em IEEE Wireless Communications and Networking Conference}, pp.
  42--47, Atlanta, GA, Apr. March.

\end{thebibliography}

\end{document}